\documentclass{article}
\usepackage{amsmath,graphicx,mlspconf}

\usepackage{amssymb}

\usepackage{url}

\graphicspath{{./figures/}}

\newcommand {\R} {{\mathbb{R}}}
\newcommand {\Z} {{\mathbb{Z}}}

\newcommand {\om} {{\omega}}

\newcommand {\lau} {{\lambda_1}}
\newcommand {\lad} {{\lambda_2}}

\def\argmin[#1]{{\operatorname*{arg\min}_{#1}}}

\newcommand {\wavu} {\psi}
\newcommand {\lowu} {\phi}
\newcommand {\indu} {\lau}
\newcommand {\indd} {\lad}

\newcommand {\Su} {S_1}

\newcommand {\Sd} {S_2}

\newcommand {\convu} {\ast}
\newcommand {\convd} {\ast}
\newcommand {\wavt} {\wavu}

\newcommand {\indt} {\alpha}

\newcommand {\wavf} {\wavu}

\newcommand {\indf} {\beta}

\newcommand {\Wav} {\Psi}

\newcommand {\sgn} {\operatorname{sgn}}

\copyrightnotice{978-1-4673-7454-5/15/\$31.00 {\copyright}2015 IEEE}

\toappear{2015 IEEE International Workshop on Machine Learning for Signal Processing, Sept.\ 17--20, 2015, Boston, USA}

\title{Joint Time-Frequency Scattering for Audio Classification}

\twoauthors
{Joakim And\'en}
{PACM\\
Princeton University, Princeton, NJ, USA\\
janden@math.princeton.edu}
{Vincent Lostanlen, St\'ephane Mallat \sthanks{This work is supported by the ERC InvariantClass 320959.}}
{D\'epartement d'Informatique\\
\'Ecole normale sup\'erieure, Paris, France\\
}

\begin{document}

\maketitle
\begin{abstract}
We introduce the joint time-frequency scattering transform, a time shift
invariant descriptor of time-frequency structure for audio classification. It
is obtained by applying a two-dimensional wavelet transform in time and
log-frequency to a time-frequency wavelet scalogram. We show that this
descriptor successfully characterizes complex time-frequency phenomena such as
time-varying filters and frequency modulated excitations.  State-of-the-art
results are achieved for signal reconstruction and phone segment classification
on the TIMIT dataset.
\end{abstract}
\section{Introduction}
\label{sec:intro}
Signal representations for classification need to capture discriminative
information from signals while remaining invariant to irrelevant variability.
This allows accurate classifiers to be trained using a limited set of labeled
examples. In audio classification, classes are often invariant to time shifts,
making time shift invariant descriptors particularly useful.

Mel-frequency spectral coefficients are time-frequency descriptors invariant to
time shifts up to $25~\mathrm{ms}$ and form the basis for the popular
mel-frequency cepstral coefficients (MFCCs) \cite{davis-mermelstein}. These can
be seen as the time-averaging of a wavelet scalogram, which is obtained by
constant-Q wavelet filtering followed by a complex modulus \cite{dss}. The time
scattering transform refines this while maintaining invariance by further
decomposing each frequency band in the wavelet scalogram using another
scalogram \cite{dss,stephane}. The result can be seen as the output of
a multilayer convolutional network \cite{stephane}. Classification experiments
have demonstrated the importance of this second layer, which captures amplitude
modulation \cite{dss}. Yet because it decomposes each frequency band
separately, it fails to capture more complex time-frequency structure such as
time-varying filters and frequency modulation, which are important in many
classification tasks.

Section \ref{sec:scattering} introduces the joint time-frequency scattering
transform which extends the time scattering by replacing the second-layer
wavelet transform in time with a two-dimensional wavelet transform in time and
log-frequency. This is inspired by the neurophysiological models of S. Shamma,
where the scalogram-like output of the cochlea is decomposed using
two-dimensional Gabor filters \cite{shihab}. Section \ref{sec:prop} shows that
joint time-frequency scattering better captures the time-frequency structure of
the scalogram by adequately characterizing time-varying filters and frequency
modulation. This is illustrated in Section \ref{sec:recon}, which presents
signal reconstruction results from joint time-frequency scattering coefficients
that are comparable to state-of-the-art algorithms and superior to time
scattering reconstruction. In Section \ref{sec:classif}, the joint
time-frequency scattering transform is shown to achieve state-of-the-art
performance for phone segment classification on the TIMIT dataset,
demonstrating the importance of properly describing time-frequency structure.
All figures and numerical results are reproducible using a MATLAB software
package available at
%\textsf{\href{http://www.di.ens.fr/data/scattering/}{http://www.di.ens.fr/data/scattering/}}.
\url{http://www.di.ens.fr/data/scattering/}.

\section{Joint time-frequency scattering}
\label{sec:scattering}
The wavelet scalogram of a signal represents time-frequency structure through a
wavelet decomposition, which filters a signal using a constant-Q wavelet filter
bank. A time scattering transform captures the temporal evolution of each
frequency band by another set of wavelet convolutions in time. It does not
fully capture the time-frequency structure of the scalogram since it neglects
correlation across frequencies. The joint time-frequency scattering remedies
this by replacing the one-dimensional wavelet transform in time with a
two-dimensional wavelet transform in time and log-frequency.

We denote the Fourier transform of a signal $x(t)$ by $\widehat{x}(\om) = \int
x(u)e^{-i\om u}\,du$. An analytic mother wavelet is a complex filter $\wavu(t)$
whose Fourier transform $\widehat{\wavu}(\omega)$ is concentrated over the
frequency interval $[1-2^{1/2Q},1+2^{1/2Q}]$. Dilations of this mother wavelet
defines a family of filters centered at frequencies $\lambda_1 = 2^{j_1/Q}$ for
$j_1 \in \Z$, given by
\begin{equation}
\label{eq:wavelet-def}
	\wavu_\indu(t) = \indu\wavu(\indu t)\quad\Longrightarrow\quad
	\widehat{\wavu}_\indu(\om) = \widehat{\wavu}(\indu^{-1}\om)~.
\end{equation}
Letting $\log u$ denote the base-two logarithm of $u$, we observe that $\log
\lambda_1 = j_1/Q$ samples each octave uniformly with $Q$ wavelets. The
temporal support of $\psi_{\indu}$ is approximately $2\pi Q/\indu$, so to
ensure that the support does not exceed some fixed window size $T$, we define
$\wavu_\indu$ using \eqref{eq:wavelet-def} only when $\indu \geq 2\pi Q/T$. The
low-frequency interval $[0, 2\pi Q/T]$ is covered by linearly spaced filters of
constant bandwidth $2\pi/T$. However, to simplify explanations, we shall
treat all filters as dilations of $\wavu$.

\begin{figure}[t]
	\begin{center}
		\setlength{\unitlength}{1cm}
		\begin{picture}(8.5,4.15)
		\put(1.7,3.75){(a)}
		\put(0.2,0){\includegraphics[width=3.5cm]{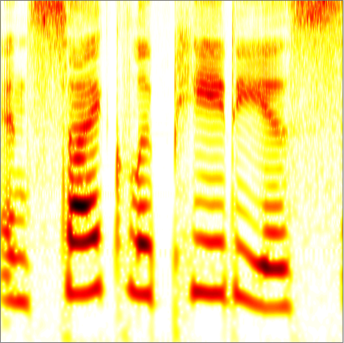}}
		\put(0.2,0){\vector(1,0){3.8}}
		\put(3.85,0.15){\small $t$}
		\put(0.2,0){\vector(0,1){3.8}}
		\put(0.30,3.65){\small $\log \indu$}
		\put(6.0,3.75){(b)}
		\put(4.5,0){\includegraphics[width=3.5cm]{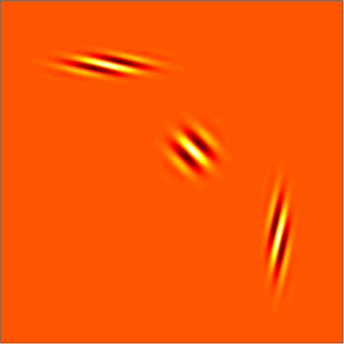}}
		\put(4.5,0){\vector(1,0){3.8}}
		\put(8.15,0.15){\small $t$}
		\put(4.5,0){\vector(0,1){3.8}}
		\put(4.6,3.65){\small $\log \indu$}
		\end{picture}
	\end{center}
	\caption{\label{fig:sample-scal} (a) Scalogram of a woman saying the
word ``encyclopedias'' with $Q = 8$ and $T = 32~\mathrm{ms}$. (b) Three examples
of real parts of wavelets $\Wav_\indd$ in the $(t,\log \indu)$-plane.}
\end{figure}

The wavelet transform convolves a signal $x$ with a wavelet filter bank. Its
complex modulus is the wavelet scalogram
\begin{equation}
	x_1(t, \log \indu) = |x \convu \wavu_\indu (t)|~,
	\quad \mbox{for~all~} \indu > 0~,
\end{equation}
an image uniformly sampled in $t$ and $\log \indu$. Here $x_1(t,\log \indu)$
represents time-frequency intensity in the interval of duration $2\pi Q/\indu$
centered at $t$ and the frequency band of bandwidth $\indu/Q$ centered at
$\indu$. Figure \ref{fig:sample-scal}(a) shows a sample scalogram.

While a rich descriptor of time-frequency structure, the scalogram is not time
shift invariant. The scattering transform ensures invariance to time shifts
smaller than $T$ by time-averaging with a low-pass filter $\phi_T$ of support
$T$, giving
\begin{equation}
\label{firnsfds}
	S_1 x(t, \log \indu) = x_1(\cdot, \log \indu) \convu \phi_T (t)~,
\end{equation}
known as first-order scattering coefficients. These approximate mel-frequency
spectral coefficients \cite{dss}.

To recover the high frequencies lost when averaging by $\phi_T$ in
\eqref{firnsfds}, $x_1$ is convolved with a second set
of wavelets $\wavu_\indd$. Computing the modulus gives
\begin{equation}
	x_2(t, \log \indu, \log \indd) =
	|x_1(\cdot, \log \indu) \convu \wavu_\indd(t)|~.
\end{equation}
As before, averaging in time creates invariance and yields
\begin{equation}
\label{2ndfoisdf}
	S_2 x(t, \log \indu, \log \indd) =
	x_2(\cdot, \log \indu, \log \indd) \convu \phi_T (t)~.
\end{equation}
These are called second-order time scattering coefficients. They supplement the
first order (and by extension mel-frequency spectral coefficients) by capturing
the temporal variability of the scalogram \cite{stephane}. Higher-order
coefficients can also be computed by repeating the same procedure.

A representation similar to second-order time scattering is the constant-Q
modulation spectrogram, which computes the spectrogram of each frequency band
and averages using a constant-Q scale \cite{thompson-atlas}. The cascade
structure of alternating convolutions and modulus nonlinearities is also shared
by convolutional neural networks, which enjoy significant success in many
classification tasks \cite{cnn,lee}.

In addition to time shift invariance, the scattering transform is also stable
to time warping due to the constant-Q structure of the wavelets
\cite{stephane}. This is useful in audio classification where small
deformations do not alter class membership.

In many audio classification tasks, such as speech recognition, classes are
invariant to frequency transposition. In this case classifiers benefit from
transposition-invariant descriptors. The time scattering transform is made
invariant to transposition by computing a frequency scattering transform along
$\log \indu$, improving classification accuracy for such tasks \cite{dss}.

While the time scattering transform successfully describes the average spectral
envelope and amplitude modulation of a signal \cite{dss}, it decomposes and
averages each frequency band separately and so cannot capture the
relationship between local temporal structure across frequency. Hence it does
not adequately characterize more complex time-frequency phenomena, such
as time-varying filters and frequency modulation.

To capture the variability of the scalogram across both time and log-frequency,
we replace the one-dimensional wavelet transform in time with a two-dimensional
wavelet transform in time and log-frequency. This follows the cortical model
introduced by S. Shamma, where a sound is decomposed by the cochlea into a
wavelet scalogram which is then convolved by two-dimensional Gabor filters in
the auditory cortex \cite{shihab}. Representations based on this cortical model
have performed well in audio classification \cite{mesgarani,kleinschmidt}, but
often lack a mathematical justification.

Let us define the two-dimensional wavelet
\begin{equation}
	\Wav_\indd(t, \log \indu) =
	\wavt_\indt(t)\, \wavf_\indf(\log \indu)~,
\end{equation}
where $\indd = (\indt, \indf)$ for $\indt \ge 0$ and $\indf \in \R$. The time
wavelet $\wavu_\indt (t)$ is calculated with a dilation by $\indt^{-1}$ for
$\indt \ge 2\pi/T$ as in \eqref{eq:wavelet-def}, giving a Fourier transform
centered at $\indt$. For these wavelets, $Q = 1$, although the notation remains
the same. Similarly, we abuse notation and define the log-frequency wavelet by
dilating a mother wavelet $\wavf$ to get
\begin{equation}
	\wavf_\indf (\log \indu) = \indf\, \wavf(\indf \log \indu)~.
\end{equation}
The identity of the wavelet will be clear from context.

The Fourier transform of $\wavf_\indf$ is centered at the frequency $\indf$.
We shall refer to this ``frequency'' parameter $\indf$ associated with the
log-frequency variable $\log \indu$ as a ``quefrency,'' with units of
cycles per octave. Note that this is different from the standard quefrency,
which is measured in seconds.

Since the two-dimensional Fourier transform of $\Wav_\indd$ is centered at
$(\indt, \indf)$, it oscillates along the slope $\indf/\indt$. Its support
in time and log-frequency is $2\pi/\indt$ by $2\pi/\indf$. Sample wavelets are
shown in Figure \ref{fig:sample-scal}(b). To ensure invertibility of the
wavelet transform, the Fourier transforms of $\Wav_\indd$ must
cover a half-plane, hence the requirement that $\indf$ take negative values.
The sign of $\indf$ determines the direction of oscillation.

The wavelet transform of $x_1$ is calculated through a two-dimensional
convolution with $\Wav_\indd$. Taking the modulus gives
\begin{equation}
	x_2(t, \log \indu, \log \indd) =
	|x_1 \convd \Wav_\indd(t, \log \indu)|~,
\end{equation}
where $\log \indd = (\log \indt, \log |\indf|, \sgn \indf)$. Similarly to
\eqref{2ndfoisdf}, second-order time-frequency scattering coefficients are
computed by time-averaging, which yields
\begin{equation}
	\Sd x(t,\log \indu, \log \indd) = x_2(\cdot, \log \indu, \log \indd) 
	\convd \phi_T(t)~.
\end{equation}
Higher-order coefficients are obtained as before by repeating the above
process. In contrast to the time scattering transform, the joint descriptor
successfully captures the two-dimensional structure of the scalogram at time
scales below $T$.

To obtain frequency transposition invariance, it would suffice to average both
$\Su x$ and $\Sd x$ along $\log \indu$ using a frequency window. However, the
amount of invariance needed may differ between classes. Since the invariant is
created through a linear mapping -- averaging along $\log \indu$ -- a
discriminative linear classifier can learn the proper amount of invariance for
each class \cite{dss}.

Just as time scattering is invariant to deformation in time, the
two-dimensional wavelet decomposition ensures that the frequency-averaged joint
scattering transform is invariant to deformation of the scalogram in time and
log-frequency. This is useful for many audio classification tasks, where
classes are often invariant under small deformations of the scalogram.

\section{Scattering time-frequency structure}
\label{sec:prop}
We apply the joint time-frequency scattering transform to two signal models: a
fixed excitation convolved with a time-varying filter and an unfiltered
frequency-modulated excitation. Both represent non-separable time-frequency
structure and are insufficiently captured by the time scattering transform but
well characterized by joint time-frequency scattering. These models do not
model more advanced structures such as polyphony and inharmonicity, but
allow us to explore the basic properties of the joint scattering transform.

\subsection{Time-varying filter}
\label{sec:filt}

Let us consider a harmonic excitation
\begin{equation}
	e(t) = \frac{2\pi}{\xi} \sum_{n} \delta\left(t-\frac{2\pi n}{\xi}\right) = \sum_k e^{i k \xi t}
\end{equation}
of pitch $\xi$. The signal is then given by applying a time-varying filter
$h(t,u)$ to $e(t)$, defined as
\begin{equation}
	x(t) = \int e(t-u) h(t,u)\,du~.
\end{equation}
Parseval's theorem now gives
\begin{equation}
	\label{eq:timevarying}
	x(t) = \frac{1}{2\pi} \int \widehat{e}(\om) \widehat{h}(t,\om) e^{i\om t}\,d\om~,
\end{equation}
where $\widehat{h}(t,\om)$ is the Fourier transform of $h(t,u)$ along $u$. Thus
$x(t)$ is the inverse Fourier transform of $\widehat{e}(\om)$ multiplied by a
time-varying transfer function $\widehat{h}(t,\om)$. These transforms are also
known as pseudo-differential operators.

Time-varying filters appear in many audio signals and carry important
information. For example, during speech production the vocal tract is deformed
to produce a sequence of phones. This produces amplitude modulation, but also
shifts formants in the spectral envelope, which can be modeled by a
time-varying filter. Similarly, much of the instrument-specific information in
a musical note is contained in the attack, which is often characterized by a
changing spectral envelope. For these reasons, it is important for an audio
descriptor to adequately capture time-varying filters.

For a suitable choice of $\indu$ we can show that
\begin{equation}
	\label{eq:filt-first}
	\Su x (t, \log \indu) \approx |\widehat{\wavu}_\indu(k\xi)|~|\widehat{h}(\cdot, \indu)|
\convu \lowu_T(t)~,
\end{equation}
where $k = \lfloor \indu/\xi \rceil$ is the index of the partial closest to
$\indu$, while for small enough quefrencies $|\indf|$
\begin{equation}
	\label{eq:filt-second}
	\def\arraystretch{1.4}
	\begin{array}{l}
		\Sd x (t, \log \indu, \log \indd) \\
		\quad \approx C \xi^{-1}|\widetilde{h} \convd \Wav_\indd(\cdot, \log \indu)| \convu \lowu_T(t)
	\end{array}~,
\end{equation}
where $C$ does not depend on $x$. Here $\widetilde{h}(t,\log \om)$ is a
weighted and log-scaled version of $|\widehat{h}(t,\om)|$ given by
$\widetilde{h}(t, \log \om) = \om |\widehat{h}(t, \om)|$. First-order
coefficients thus provide the time-averaged amplitude of $\widehat{h}$ sampled
at the partials $k\xi$ since $|\widehat{\wavu}_\indu(k\xi)|$ is non-negligible
only for $\lau \approx k\xi$. Furthermore, the second order approximates the
two-dimensional scattering coefficients of the modified filter
transfer function $\widetilde{h}$, capturing its time-frequency structure.

In contrast, the time scattering transform only characterizes separable
time-varying filters $h$ that can be written as the product of an amplitude
modulation in time and a fixed filter. In this case the model reduces to the
amplitude-modulated, filtered excitation considered in \cite{dss}. Time
scattering and joint time-frequency scattering thus differ in that the latter
captures the non-separable structure of $h$ while the former only describes its
separable structure.

To justify \eqref{eq:filt-first} and \eqref{eq:filt-second}, we proceed as in
\cite{dss}, convolving \eqref{eq:timevarying} with $\wavu_\indu$ and taking the
modulus to obtain
\begin{equation}
	x_1(t, \log_2 \indu) \approx |\widehat{h}(t,\indu)| \sum_k |\widehat{\wavu}_\indu(k\xi)|~,
\end{equation}
for $\widehat{h}(t,\om)$ smooth enough and $\lau/Q < \xi$. In this case at most
one partial $k\xi \approx \lau$ is found in the support of the wavelet so the
sum only contains one non-negligible term when $k = \lfloor \indu/\xi \rceil$.
Averaging in time yields \eqref{eq:filt-first}. Furthermore, we note that as a
function of $\log \indu$, the sequence of partials $\sum_k
|\widehat{\wavu}_\indu(k\xi)|$ can be approximated at large scale by $C \indu
\xi^{-1}$. For small $|\indf|$, $\wavf_\indf$ is very regular in $\log \indu$.
If $|\widehat{h}(t,\om)|$ is also smooth enough along $\om$, we can therefore replace the
sum of partials by $C \indu \xi^{-1}$ when convolving $x_1$ with $\Wav_\indd$.
Rewriting the convolution using $\widetilde{h}$ then yields
\begin{equation}
	x_1 \convd \Wav_\indd(t,\log \indu) \approx C \xi^{-1} \widetilde{h} \convd \Wav_\indd(t, \log \indu)~.
\end{equation}
Taking the modulus and averaging then gives \eqref{eq:filt-second}.

\subsection{Frequency modulation}
\label{sec:fm}

We now consider an excitation of varying pitch
\begin{equation}
	\label{eq:fm}
	x(t) = \sum_k e^{ik\theta(t)}~.
\end{equation}
At time $t$, $x$ has instantaneous pitch $\theta'(t)$ and relative pitch
variation $\theta''(t)/\theta'(t)$. This carries important information in many
sounds, such as tonal speech, bioacoustic signals, and music (e.g. for vibratos
and glissandi). A good audio descriptor should therefore adequately describe
such pitch changes.

For appropriate $\indu$ and $T$, we can show that
\begin{equation}
	\label{eq:fm-first}
	\Su x(t, \log \indu) \approx |\widehat{\wavu}_\indu(k\theta'(\cdot))| \convu \lowu_T(t)~,
\end{equation}
where $k = \lfloor \indu/\theta'(t) \rceil$ as before. Furthermore, for $|\beta|$ large,
\begin{equation}
	\label{eq:fm-second}
	\def\arraystretch{1.4}
	\begin{array}{l}
		\Sd x(t, \log \indu, \log \indd) \\
		\quad \approx
		C \left( S_1 x(t, \cdot) \convu \phi_{2\pi/\beta}(\log \indu) \right)
		\left|\widehat{\wavu}\left(-\frac{\indf \theta''(t)}
			{\indt \theta'(t)}\right)\right|
	\end{array}~,
\end{equation}
where $C$ is independent of $x$.

While first-order joint scattering coefficients provide an average of the
instantaneous pitch $\theta'(t)$ over the interval of duration $T$, the second
order describes the rate of pitch variation $\theta''(t)/\theta'(t)$. Indeed,
for fixed $t$ and $\indu$, $\Sd x$ is maximized along the line $\indt/\indf =
-\theta''(t)/\theta'(t)$, and so captures this frequency modulation structure.
The time scattering transform, in contrast, only captures the bumps in each
frequency band induced by the varying pitch, ignoring its frequency structure.

To see why \eqref{eq:fm-first} and \eqref{eq:fm-second} hold, we linearize
$\theta(t)$ over the support of $\wavu_\indu$ when decomposing \eqref{eq:fm},
which gives
\begin{equation}
	x_1(t,\log \indu) \approx |\widehat{\wavu}_\indu(k\theta'(t))|~,
\end{equation}
provided that $\indu/Q < \|\theta'\|_\infty$. As before, only the partial $k =
\lfloor \indu/\theta'(t) \rceil$ is contained in the frequency support of
$\wavu_\indu$. Averaging in time gives \eqref{eq:fm-first}. Each partial traces
a curve along $\lau = k\theta'(t)$, so locally the scalogram $x_1$ can be approximated
by sliding Dirac functions $C\delta(\log \indu - \log k\theta'(t))$ for some
$C$. Convolving $x_1$ along $\log \indu$ with $\wavf_\indf$ for $|\indf|$ large
enough to capture only one line gives $C\wavf_\indf(\log \indu - \log
k\theta'(t))$. For a fixed $\indu$, this is a complex exponential of
instantaneous frequency $-\indf\theta''(t)/\theta'(t)$ multiplied by an
envelope. Convolving this in time with a wavelet $\wavt_\indt$ on whose support
the envelope is approximately constant then gives
\begin{equation}
	\def\arraystretch{1.4}
	\begin{array}{l}
		x_1 \convu \Wav_\indd(t,\log \indu) \\
		\quad \approx C\wavf_\indf\left(\log \frac{\indu}{k\theta'(t)}\right)\,
		\widehat{\wavu}\left(-\frac{\beta \theta''(t)}{\alpha\theta'(t)}\right)
	\end{array}~,
\end{equation}
Taking the modulus, we can replace $|\wavf_\indf|$ with the low-pass filter
$\lowu_{2\pi/\indf}$. Assuming that $\theta''(t)/\theta'(t)$ is almost constant
over an interval of duration $T$, averaging gives \eqref{eq:fm-second}.

We note that the time-varying filter and frequency modulation models in
\eqref{eq:timevarying} and \eqref{eq:fm} are complementary. For small quefrencies
$|\beta|$, the joint scattering coefficients capture time-frequency structure
over large frequency intervals, which is given by time-varying filters. Larger
$|\beta|$ describe more localized behavior in log-frequency, like frequency
modulation. This scale separation allows the joint scattering transform to
simultaneously characterize both types of structures.

\section{Time-shift invariant reconstruction}
\label{sec:recon}

After having analyzed a given signal $x$ with a scattering transform,
synthesizing a new signal $y$ from the invariant
coefficients $S_1 x$ and $S_2 x$ highlights what information
is captured in the representation \textemdash{} and, conversely,
what is lost. In this section, we use a backpropagation algorithm
on stationary audio textures to qualitatively compare the
joint scattering transform with other architectures.

The reconstruction $y$ is first initialized with random noise, and then
iteratively updated to converge to a local minimum of the functional
\begin{equation}
\Vert Sx-Sy \Vert^{2} =
\Vert S_1 x - S_1 y \Vert^2 + \Vert S_2 x - S_2 y \Vert^2
\end{equation}
with respect to $y$.  Since the forward computation of scattering coefficients
consists of an alternated sequence of linear operators (wavelet convolutions)
and modulus nonlinearities, the chain rule for gradient backpropagation yields
a sequence of closed-form derivatives in the reverse order. The modulus
nonlinearities are backpropagated by applying $|z(t)|' = {\rm Real}(z'(t)
\,|z(t)| /z(t))$.  In turn, the backpropagation of the wavelet transforms
consists of convolving each frequency band by the complex conjugate of the
corresponding wavelet and summing across bands \cite{bruna-texture}.

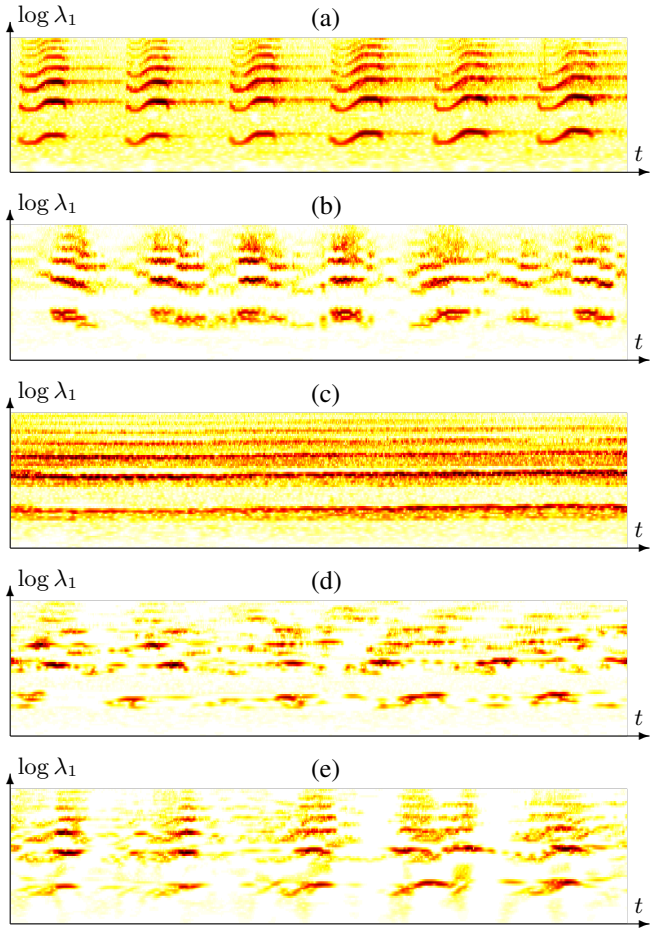
\begin{figure}[t]
	\begin{center}
		\setlength{\unitlength}{1cm}
		\begin{picture}(8.5,12.1)
		\put(4,11.65){(a)}
		\put(0,9.7){\includegraphics[height=1.8cm,width=8.2cm]{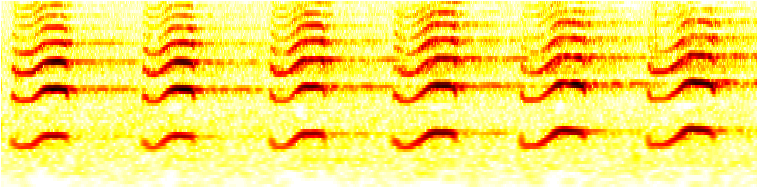}}
		\put(0,9.7){\vector(1,0){8.5}}
		\put(8.3,9.85){\small $t$}
		\put(0,9.7){\vector(0,1){2}}
		\put(0.1,11.7){\small $\log \indu$}
		
		\put(4,9.15){(b)}
		\put(0,7.2){\includegraphics[height=1.8cm,width=8.2cm]{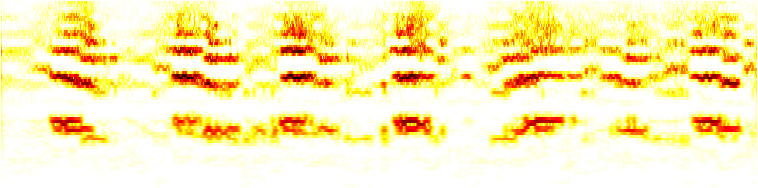}}
		\put(0,7.2){\vector(1,0){8.5}}
		\put(8.3,7.35){\small $t$}
		\put(0,7.2){\vector(0,1){2}}
		\put(0.1,9.2){\small $\log \indu$}
		
		\put(4,6.65){(c)}
		\put(0,4.7){\includegraphics[height=1.8cm,width=8.2cm]{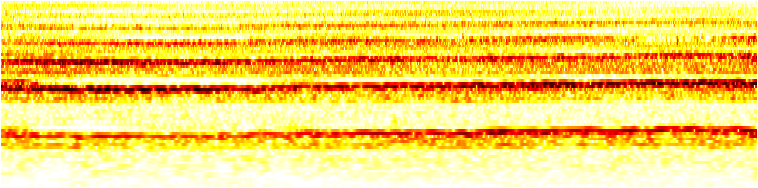}}
		\put(0,4.7){\vector(1,0){8.5}}
		\put(8.3,4.85){\small $t$}
		\put(0,4.7){\vector(0,1){2}}
		\put(0.1,6.7){\small $\log \indu$}
		
		\put(4,4.15){(d)}
		\put(0,2.2){\includegraphics[height=1.8cm,width=8.2cm]{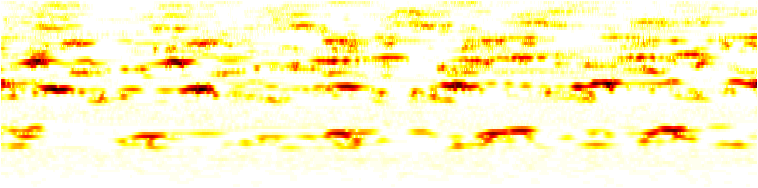}}
		\put(0,2.2){\vector(1,0){8.5}}
		\put(8.3,2.35){\small $t$}
		\put(0,2.2){\vector(0,1){2}}
		\put(0.1,4.2){\small $\log \indu$}
		
		\put(4,1.65){(e)}
		\put(0,-0.3){\includegraphics[height=1.8cm,width=8.2cm]{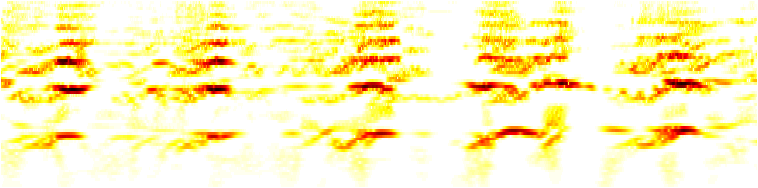}}
		\put(0,-0.3){\vector(1,0){8.5}}
		\put(8.3,-0.15){\small $t$}
		\put(0,-0.3){\vector(0,1){2}}
		\put(0.1,1.7){\small $\log \indu$}
		\end{picture}
	\end{center}
	\protect\caption{
\label{fig:bird-synthesis}
Reconstructed bird calls from time-invariant coefficients.
Top to bottom:
(a) original $x(t)$,
(b) from McDermott and Simoncelli representation \cite{mcdermott},
(c) from first-order scattering,
(d) from first- and second-order time scattering,
(e) from first- and second-order joint time-frequency scattering.}
\end{figure}

To illustrate, we have synthesized a bird song recording using different
scattering transforms. Here $T=375\,\mathrm{ms}$ and is of the order of three
bird calls (see Figure \ref{fig:bird-synthesis}(a)). First-order coefficients
$S_{1}x$ yield the reconstruction in Figure \ref{fig:bird-synthesis}(c). This
fits the averaged mel-frequency spectrum of the target sound. Although this is
sufficient when $x$ is the realization of a Gaussian process, it does not
convey the typical intermittency in natural sounds. This is partly mitigated by
adding second-order coefficients, giving the reconstruction in Figure
\ref{fig:bird-synthesis}(d), since these encode the amplitude modulation
spectra in each acoustic subband.  However, these spectra are not synchronized
across subbands, so time scattering tends to synthesize auditory textures made
of decorrelated impulses. In contrast, we observe that the reconstruction from
joint scattering coefficients in Figure \ref{fig:bird-synthesis}(e) is able to
capture coherent structures in the time-frequency plane, such as joint
modulations in amplitude and frequency.  Notably, because of their chirping
structure, bird calls are better synthesized with joint scattering. Indeed,
recalling \eqref{eq:fm-second}, chirps are represented with few nonzero
coefficients in the basis of joint time-frequency wavelets. We believe that
audio re-synthesis is greatly helped by this gain in sparsity. More experiments
are available at
%\textsf{\href{http://www.di.ens.fr/data/scattering/audio/}{http://www.di.ens.fr/data/scattering/audio/}}.
\url{http://www.di.ens.fr/data/scattering/audio/}.

McDermott and Simoncelli \cite{mcdermott} have built an audio texture synthesis
algorithm based on a scattering-like transform along time, of which they
compute cross-correlation statistics across $\lambda_1$ and across $\lambda_2$,
as well as marginal moments (variance and skewness). Their representation is
also able to synchronize frequency bands and recover amplitude modulation.
Nevertheless, asymmetry in frequency modulation is lost. Indeed, while all bird
calls from the original recording have an ascending instantaneous frequency,
some of the chirps reconstructed with their method descend instead.  Moreover,
the higher-order statistics on which they rely are unstable to deformations and
hence not suitable for classification purposes. In this section, we have shown
that joint scattering may achieve comparable or better quality in audio
re-synthesis, yet with only using stable features.

On the negative side, it must be noted that joint scattering is insufficient to
capture temporal changes in harmonic structure.  Indeed, partial tones which
are several octaves apart are not likely to be correctly in tune \textemdash{}
a limitation that we shall specifically address as a future work.

\section{Classification}
\label{sec:classif}

We evaluate the performance of the joint time-frequency scattering
representation on phone segment classification using the TIMIT dataset
\cite{timit}. The corpus consists of $6300$ phrases, each of which has its
constituent phone segments labeled with its position, duration, and identity.
Given a position and duration, we want to identify the phone contained in the
segment. This task is easier than the problem of continuous speech recognition,
but provides a straightforward framework when evaluating signal representations
for speech.

We follow the same setup as in \cite{dss}. Each phone is represented by a given
descriptor applied to a $192$-millisecond window centered on the phone along
with the phone's log-duration. A Gaussian support vector machine (SVM) is used
as a classifier through the LIBSVM library \cite{libsvm}.

The SVM is a discriminatively trained, locally linear classifier. This means
that, given enough training data, an SVM can learn the amount of averaging
needed along $\log \indu$ to gain the desired invariance \cite{dss}. We
therefore present results for scattering transforms without averaging along
$\log \indu$.

Table \ref{table:results} shows the results of the classification task. MFCCs
calculated over the segment with a window size of $32~\mathrm{ms}$ and
concatenated to yield a single feature vector provide a baseline error rate of
$18.3\%$. The non-scattering state of the art achieves $16.7\%$ and is obtained
using a committee-based hierarchical discriminative classifier on MFCC
descriptors \cite{chang-glass}. A convolutional network classifier applied to
the log-scalogram with learned filters obtains $19.7\%$ \cite{lee}.

\begin{table}
	\begin{center}
	\begin{tabular}{|l|c|}
		\hline
		Representation & Error rate (\%) \\
		\hline
		MFCCs & $18.3$ \\ % phones_261
		State of the art (excl. scattering) \cite{chang-glass} & $16.7$ \\
		\hline
		Time Scattering & $17.3$ \\ % phones_253, 1*T_s, renorm_eps = 2^0
		Time Scattering + Freq. Scattering & $16.1$ \\ % phones_274, dur_mult = 2^4
		Joint Time-Freq. Scattering & $15.8$ \\ % phones_273, dur_mult = 2^4
		\hline
	\end{tabular}
	\end{center}
	\caption{\label{table:results} Error rates in percent for the phone segment
classification task. MFCCs and scattering transforms are computed with $T =
32~\mathrm{ms}$ and $Q = 8$.}
\end{table}

The time scattering transform is computed with $T = 32~\mathrm{ms}$ and $Q = 8$
up to the second order. As in previous experiments, we compute the logarithm of
the scattering \cite{dss}. Since it better captures amplitude modulation,
results improve with respect to MFCCs, achieving an error of $17.3\%$.

Applying an unaveraged frequential scattering transform along $\log \indu$ up
to a scale of $K = 4$ octaves and computing the logarithm yields an error rate
of $16.1\%$. As discussed earlier, transposition invariance counters speaker
variability, and so improves performance. However, the frequency scattering is
computed along $\log \indu$ of a time scattering transform which has been
averaged in time, so its discriminability also suffers from not capturing
local correlations across frequencies.

Computing the joint time-frequency scattering transform for $K = 4$ octaves
yields an error of $15.8\%$, an improvement compared to the time scattering
transform with scattering along log-frequency. This illustrates the importance
of the complex time-frequency structure that is captured by the joint
scattering transform, and can be partly explained by the fact that the onset of
many phones is characterized by rapid changes in formants, which can be modeled
by time-varying filters. As we saw earlier, these are better described by
time-frequency scattering compared to time scattering. However, the small
window size $T$ limits the loss of time-frequency structure in the time
scattering transform. We therefore expect a greater improvement for tasks
involving larger time scales.

The previous state of the art was obtained at $15.9\%$ using a scattering
transform with multiple Q factors \cite{dss}. This more ad hoc descriptor has
many similarities with the joint scattering transform, but is difficult to
study analytically.

\section{Conclusion}
\label{sec:conclusion}

We introduced the joint time-frequency scattering transform, which is a
time-shift invariant representation stable to time-frequency warping. This
representation characterizes time-varying filters and frequency modulation.
Reconstruction experiments show how it successfully captures complex
time-frequency structures of locally stationary signals. Finally, phone segment
classification results demonstrate the value of adequately representing these
structures for classification.

\bibliographystyle{IEEEbib}
\bibliography{refs}

\end{document}